\documentclass[11pt]{article}
\usepackage{amsmath,amsthm,amssymb,cite,enumerate}
\usepackage[a4paper,left=0.5in,right=1in]{geometry}
\usepackage[usenames]{color}
\usepackage{graphicx}
\usepackage{epsfig}
\usepackage{dcolumn}
\usepackage{bm}
\usepackage{authblk} 

\begin{document}

\def \d {{\rm d}}

\def \bm {\mbox{\boldmath{$m$}}}

\def \bF {\mbox{\boldmath{$F$}}}
\def \bV {\mbox{\boldmath{$V$}}}
\def \bff {\mbox{\boldmath{$f$}}}
\def \bT {\mbox{\boldmath{$T$}}}
\def \bk {\mbox{\boldmath{$k$}}}
\def \bl {\mbox{\boldmath{$\ell$}}}
\def \bn {\mbox{\boldmath{$n$}}}
\def \bbm {\mbox{\boldmath{$m$}}}
\def \tbbm {\mbox{\boldmath{$\bar m$}}}
\def \l {\ell}

\newcommand*\bg{\ensuremath{\boldsymbol{g}}}
\newcommand*\bE{\ensuremath{\boldsymbol{E}}}
\newcommand*\bh{\ensuremath{\boldsymbol{h}}}
\newcommand*\bR{\ensuremath{\boldsymbol{R}}}
\newcommand*\bu{\ensuremath{\boldsymbol{u}}}
\newcommand*\bA{\ensuremath{\boldsymbol{A}}}
\newcommand*\bep{\ensuremath{\boldsymbol{\varepsilon}}}
\newcommand*\bPsi{\ensuremath{\boldsymbol{\Psi}}}

\def \T {\bigtriangleup}
\newcommand{\msub}[2]{m^{(#1)}_{#2}}
\newcommand{\msup}[2]{m_{(#1)}^{#2}}

\newcommand{\be}{\begin{equation}}
\newcommand{\ee}{\end{equation}}

\newcommand{\beq}{\begin{eqnarray}}
\newcommand{\eeq}{\end{eqnarray}}
\newcommand{\pa}{\partial}
\newcommand{\pp}{{\it pp\,}-}
\newcommand{\ba}{\begin{array}}
\newcommand{\ea}{\end{array}}

\newcommand{\M}[3] {{\stackrel{#1}{M}}_{{#2}{#3}}}
\newcommand{\m}[3] {{\stackrel{\hspace{.3cm}#1}{m}}_{\!{#2}{#3}}\,}

\newcommand{\tr}{\textcolor{red}}
\newcommand{\tb}{\textcolor{blue}}
\newcommand{\tg}{\textcolor{green}}

\def\a{\alpha}
\def\g{\gamma}
\def\de{\delta}
\def\b{\beta}

\def\E{{\cal E}}
\def\B{{\cal B}}
\def\R{{\cal R}}
\def\F{{\cal F}}
\def\L{{\cal L}}
\def\F2{{F}}

\def\e{e}
\def\bb{b}

\newtheorem{theorem}{Theorem}[section] 
\newtheorem{cor}[theorem]{Corollary} 
\newtheorem{lemma}[theorem]{Lemma} 
\newtheorem{prop}[theorem]{Proposition}
\newtheorem{definition}[theorem]{Definition}
\newtheorem{remark}[theorem]{Remark}

\title{Static and radiating dyonic black holes coupled to conformally invariant electrodynamics in higher dimensions}

\author[1,2]{David Koko\v ska\thanks{david.kokoska@matfyz.cz}}
\author[2]{Marcello Ortaggio\thanks{ortaggio(at)math(dot)cas(dot)cz}}

\affil[1]{Institute of Theoretical Physics, Faculty of Mathematics and Physics, \newline
 Charles University, V Hole\v{s}ovi\v{c}k\'{a}ch 2, 180 00 Prague 8, Czech Republic}

\affil[2]{Institute of Mathematics of the Czech Academy of Sciences, \newline \v Zitn\' a 25, 115 67 Prague 1, Czech Republic}

\maketitle

\begin{abstract}
We investigate the complete family of (aligned) Robinson-Trautman spacetimes sourced by conformally invariant non-linear electrodynamics in $D$ dimensions in the presence of an arbitrary cosmological constant. After presenting general features of the solutions (which exist only in even dimensions), we discuss in more detail some particular subclasses. Static metrics contain dyonic black holes with various possible horizon geometries (K\"ahler if there is a magnetic field, including flat branes) and different asymptotics. In addition, there exist also time-dependent solutions (not possible in the $D>4$ linear theory) which may represent white hole evaporation by emission of electromagnetic radiation (or a time-reversed picture of black hole formation). For those, we comment on a quasi-local characterization of possible past horizons. 
Finally, we briefly discuss the special case of stealth solutions. In an appendix, a theory-independent result on the redundancy of the gravity part of the field equations for Robinson-Trautman spacetimes is further obtained.
\end{abstract}

\tableofcontents

\section{Introduction}

\label{sec_intro}

Non-linear classical theories of electrodynamics were originally introduced in order to cure the divergent electron's self-energy \cite{Mie12,Born33,BorInf34}. Modified theories also naturally appear as effective Lagrangians taking into account various quantum corrections (cf., e.g., the review \cite{Dunne05} and references therein) as well as low-energy limits of string theory \cite{SchSch74,Callanetal85,FraTse85}. Coupling modified electrodynamics to gravity is clearly also of interest and it is remarkable that certain non-linearities can regularize black holes \cite{AyoGar98,Bronnikov00,AyoGar00}.

In recent years, higher-dimensional scenarios have attracted increasing attention, with motivation coming from different directions, such as string theory, the AdS/CFT correspondence, and brane-world models. While higher-dimensional static black hole solutions in the Einstein-Maxwell theory were obtained several decades ago \cite{Tangherlini63}, it is a natural question to ask how their properties are modified when Einstein's gravity is coupled to more general electrodynamics. Perhaps the simplest extensions of the Maxwell Langrangian include polynomial functions of the invariant $F\equiv F_{\mu\nu}F^{\mu\nu}$ -- yet already such simple higher-order corrections make the field equations in general much more difficult to solve. Within this class of theories, the only conformally invariant action in $D$ dimensions is defined by the monomial $F^{D/4}$ \cite{HasMar07} (see also section~\ref{subsec_theory}). While for $D=4$ this reduces to the standard Maxwell action, it gives rise to non-linear equations of motion when $D>4$.\footnote{A different (and linear) conformally invariant extension of Maxwell's theory is defined in $D=2p$ dimensions by the Lagrangian density $\sqrt{-g}F_{\a_1\ldots\a_{p}}F^{\a_1\ldots\a_{p}}$, where $F_{\a_1\ldots\a_{p}}$ is a $p$-form field \cite{Teitelboim86}. See \cite{Bandosetal21} and references therein for non-linear $p$-form theories admitting conformal invariance. \label{footn_2p}} Nevertheless, the conformal character of the matter field allows for a considerable simplification of the field equations. Indeed, a simple exact solution representing an electrically charged, spherically symmetric, asymptotically flat black hole was obtained in \cite{HasMar07} (provided $D$ is a multiple of four). This result is of interest also in that it contrasts with certain no-go theorems obtained for the case of higher-dimensional black holes coupled to a conformally invariant scalar field \cite{XanDia92,Klimcik93}.

The purpose of the present paper is to study black hole solutions of the theory proposed in \cite{HasMar07} from a more general viewpoint. To this end, we will analyze systematically the class of Robinson-Trautman solutions, which are defined by the existence of an expanding, shearfree and twistfree congruence of null geodesics (see section~\ref{sec_geom} for a short summary).\footnote{We observe that four-dimensional Robinson-Trautman solutions coupled to non-linear electrodynamics have been studied in \cite{TahSvi16,Tahamtan21}. There is however no overlap with the results of our paper, since the matter field equations of the theory~\eqref{action} become linear when $D=4$.}  In four-dimensional General Relativity, such spacetimes were first constructed in \cite{RobTra60} as an example of spherical gravitational radiation, and studied more systematically in \cite{RobTra62} and in several subsequent papers by various authors (cf. \cite{Stephanibook,GriPodbook} for reviews and for a number of references). The family of solutions of \cite{RobTra60,RobTra62} comprises diverse (electro)vacuum spacetimes ranging from static Schwarzschild-like black holes of various topologies to their accelerating (C-metric) or null fluid (Vaidya metric) counterparts, as well as more general radiative solutions. An interesting feature of the latter is that at late times (in vacuum and under suitable initial conditions) they decay to the Schwarzschild (or Kottler) spacetime by emitting gravitational waves (see \cite{GriPodbook} for a review and relevant references). They also serve as exact models where various quasi-local characterizations of horizons in dynamical situations can be tested and visualized \cite{Penrose73,Tod86,Tod89}. In the charged case with $\Lambda=0$, however, linear perturbations lead to an instability, and the physical meaning of such solutions is less clear \cite{LunCho94,KozKreReu08}.

The general form of Robinson-Trautman metrics in $D$ dimensions was obtained in \cite{PodOrt06} (see \cite{RobTra83} for an earlier discussion from a more geometrical viewpoint), where vacuum solutions were also studied. Extensions to the higher-dimensional Einstein-Maxwell theory were later obtained in \cite{OrtPodZof08}. In both cases, such family of solutions turned out to be much more restricted when $D>4$, in particular it does not contain the interesting radiative spacetimes known for $D=4$.  
The motivation to consider here the theory of \cite{HasMar07} is thus twofold. On the one hand, adopting an ansatz more general than the one used in \cite{HasMar07} enables one to explore the space of static black hole solutions more systematically, including dyonic configurations, various horizon geometries and different asymptotics (and also when $D$ is not a multiple of four, under certain conditions -- cf. the following). This can possibly be of interest in the context of generalized thermodynamics (see, e.g., the recent work \cite{BokJurSmo21} and references therein for general results, and \cite{GonHasMar09} for a discussion relevant to the theory considered here).
Furthermore, by allowing for time-dependent solutions, one can analyze to what extent the negative results of \cite{PodOrt06,OrtPodZof08} can be bypassed thanks to the conformal invariance of the action for matter fields (this is in part motivated by the results of \cite{OrtPodZof15} for the $p$-form Einstein-Maxwell theory in $D=2p$ dimensions -- cf. footnote~\ref{footn_2p}). As we will show, time-dependent solutions which can be interpreted as dynamical black holes emitting (or absorbing) electromagnetic radiation do indeed exist in the theory of \cite{HasMar07}. These are also interesting from the viewpoint of dynamical ``quasi-local'' horizons.

The plan of the paper is as follows. In section~\ref{subsec_theory} we outline the basic features of the theory under consideration \cite{HasMar07}. In section~\ref{subsec_summary} we compactly summarize our main results in the case of Robinson-Trautman non-stealth solutions. After outlining their derivation in section~\ref{sec_integr}, we describe those in more detail in sections~\ref{sec_static} and \ref{sec_radiating} in the most interesting cases of static black holes and their time-dependent (radiating) extensions. Stealth solutions are briefly discussed in section~\ref{sec_stealth}. Some concluding remarks are given in section~\ref{sec_concl}. 
Appendices~\ref{app_conserv} and \ref{app_Ricci} contain some technical results useful for the derivation of the solutions in section~\ref{sec_integr}. Namely, appendix~\ref{app_conserv} presents a result on the redundancy of the gravity part of the field equations for Robinson-Trautman spacetimes which applies to a large class of diffeomorphism-invariant metric theories of gravity arbitrarily coupled to (unspecified) matter fields. Appendix~\ref{app_Ricci} contains the components of the Ricci tensor for Robinson-Trautman metrics, needed to integrate Einstein's equations.

\subsection{The theory}

\label{subsec_theory}

We consider $D$-dimensional Einstein gravity minimally coupled to a 2-form $\bF=\d\bA$ in the following theory \cite{HasMar07}
\be
  S=\int d^Dx\sqrt{-g}\left[\frac{1}{\kappa} \left( R - 2 \Lambda \right)-2\beta\F2^{D/4}\right] ,
	\label{action}
\ee
where $\kappa$ and $\beta$ are coupling constants, and 
\be
 \F2\equiv F_{\mu\nu}F^{\mu\nu} .  
 \label{F2}
\ee

Variations of \eqref{action} w.r.t. $\bg$ and $\bA$ give rise to the equations of motion \cite{HasMar07}
\beq
 & & \frac{1}{\kappa}\left(G_{\mu\nu}+\Lambda g_{\mu\nu}\right)=\beta F^{D/4-1}\left(DF_{\mu\rho}F_\nu^{\phantom{\nu}\rho}-g_{\mu\nu}F\right) , \label{Einst} \\
 & & \frac{1}{\sqrt{-g}}\pa_{\mu}\left(\sqrt{-g}F^{\frac{D}{4}-1}F^{\mu \nu} \right)=0 . \label{Max1}
\eeq
The latter can be understood as generalized Maxwell equations. For the sake of brevity, in the following we will just refer to~\eqref{Max1} as the Maxwell equations of the theory~\eqref{action}. We will also write eq.~\eqref{Einst} compactly as $E_{\mu\nu}=0$ (cf.~\eqref{grav_eq}), in order to later refer to some of its components for specific values of the indices. The RHS of \eqref{Einst} defines the energy-momentum tensor $T_{\mu\nu}$.

In addition, $\bF$ must be closed, i.e.,
\be
	F_{[\mu \nu , \rho]}=0 .
	\label{Max2} 
\ee	

Since the RHS of \eqref{Einst} is traceless, the Ricci scalar is a constant, i.e.,
\be 
	R = \frac{2D}{D-2}\Lambda .
\ee	
For later computations we observe that this allows one to write the LHS of \eqref{Einst} as $G_{\mu\nu}+\Lambda g_{\mu\nu}=R_{\mu\nu}-2\Lambda g_{\mu\nu}/(D-2)$.

Let us observe that for $D\neq4$ the theory~\eqref{action} admits {\em stealth} solutions, i.e., non-trivial electromagnetic configurations for which the energy-momentum tensor, i.e., the RHS of \eqref{Einst}, vanishes identically.\footnote{This should be contrasted with the standard linear Maxwell theory, for which non-zero fields with vanishing energy-momentum are not permitted  (in any dimension) -- cf. e.g. footnote~10 of \cite{HerOrt20_2} and references therein. Stealth fields in non-linear four-dimensional theories have been studied in \cite{Smolic18}. See \cite{Sokolowski04} for a more general discussion on matter field Lagrangians with vanishing energy-momentum.} It is easy to prove that those are precisely the configurations such that $F=0$ (see section~\ref{sec_stealth}). This clearly also ensures that the Maxwell equations~\eqref{Max1} are identically satisfied. Therefore, any closed 2-form $\bF$ provides a solution to the theory~\eqref{action} in any Einstein spacetime.

The reality of the quantity $F^\frac{D}{4}$ (which appears in the field equations) implies that $D$ must be a multiple of 4 when $F<0$. From the above discussion it also follows that the requirement of a non-negative energy density (in an orthonormal frame) $T_{\hat0\hat0}\ge0$ amounts to $\beta F^{D/4-1}\ge0$, so that (as discussed in a special case in \cite{HasMar07}): i) $\beta>0$ if $F>0$, or if $F<0$ with $D/4$ being odd; (ii) $\beta<0$ if $F<0$ with $D/4$ being even; (iii) $\beta$ can have any sign in the stealth case $F=0$. This will be assumed in the following.

\subsection{Summary of the results (non-stealth fields)}

\label{subsec_summary}

Let us assume here that the electromagnetic field is non-stealth (i.e., $F\neq0$). 
The line-element is given by
\be
  \d s^2=r^{2}h_{ij}\d x^i\d x^j{-}2\,\d u\d r-2H\d u^2 , 
	\label{ds_generic}
\ee
where Latin indices $i,j,\ldots=1,\ldots,D-2$ label the spatial coordinates $x^i$ (from now on collectively denoted simply as $x$), and the base space metric $h_{ij}=h^{1/(D-2)}(u,x)\gamma_{ij}(x)$ represents a Riemannian Einstein space of dimension $D-2$ and scalar curvature $\R=K(D-2)(D-3)$ (we denoted $h\equiv \det h_{ij}$, such that $\gamma_{ij}$ is unimodular).

The (aligned) electromagnetic field is
\be
 \bF = \dfrac{ e}{r^2}\d r \wedge\d u + \left( \dfrac{ e_{,i}}{r}-\xi_i \right)\d u \wedge \d x^i + \dfrac{1}{2}F_{ij}\d x^i \wedge \d x^j ,
 \label{F_summ}
\ee
and the metric function $H$ in \eqref{ds_generic} is defined by
\beq
 & & 2H=K+\frac{2}{D-2}\big(\ln \sqrt{h}\big)_{,u}r -\lambda r^2 - \frac{\mu}{r^{D-3}} + \frac{Q^2}{r^{D-2}}  \qquad\qquad (K=0,\pm 1) , \label{H} \\
 & & \lambda\equiv\frac{2\Lambda}{(D-2)(D-1)} , \qquad Q^2\equiv { 2\kappa \beta} F_0^{\frac{D}{4}-1}\left(\frac{b^2}{D-2} + e^2\right) . \label{Q}
\eeq

In~\eqref{F_summ}--\eqref{Q}, the quantities $e$, $\xi_i$, $F_{ij}$, $\mu$, $b$ and $F_0$ may in general depend on $(u,x)$, with ($h^{ij}$ denotes the inverse of $h_{ij}$)
\be
  F_0\equiv  b^2-2 e^2 , \qquad  b^2\equiv F_{ik} F_{jl} h^{ij} h^{kl} , \label{F_spat}
\ee
and $F=r^{-4}F_0$. The functions $e$ and $b$ characterize the strength of, respectively, the purely electric and purely magnetic parts of the electromagnetic field~\eqref{F_summ}, while $\xi_i$ its radiative (null) component (see also footnote~\ref{footn_charge} and eq.~\eqref{flux}). By the observations in section~\ref{subsec_theory} it follows that $D$ must be a multiple of 4 when $ b^2-2 e^2<0$. The singularity of $F$ at $r=0$ represents also a (timelike) curvature singularity, as can be checked by computing, e.g., the invariant $R_{\mu\nu}R^{\mu\nu}$.

Further conditions coming from the Maxwell and Einstein equations are
\beq 
 & & F_{[ij,k]}= 0 , \qquad F_{ij,u}=\xi_{i,j}-\xi_{j,i} , \label{Fij,u} \\ 
 & & F_0^{\frac{D}{4}-1}\sqrt{h}h^{ij} e_{,j}  = \Big(F_0^{\frac{D}{4}-1} \sqrt{h}  h^{ik} h^{jl} F_{kl} \Big)_{,j} , \qquad 
\Big(F_0^{\frac{D}{4}-1}\sqrt{h}h^{ij} \xi_j \Big)_{,i} =  \Big(F_0^{\frac{D}{4}-1} \sqrt{h}   e\Big)_{,u} , \label{grad_alpha}
\eeq
and
\beq
 & & \mu_{,i} =  {2\kappa \beta} D F_0^{\frac{D}{4}-1}\big(  e \xi_i - F_{ik} \xi_j h^{kj} \big) , \label{mu_i} \\
 & & (D-2)\mu_{,u} = -(D-1) \mu \big( \ln \sqrt{h} \big)_{,u}-  {2\kappa D \beta}  F_0^{\frac{D}{4}-1}h^{ij}\xi_i \xi_j  \qquad (D>4) . \label{mu_u}
\eeq
Note that none of the above equations contain $\Lambda$.

Finally, when $\bF$ contains a non-zero magnetic component $F_{ij}\neq0$ ($\Leftrightarrow  b^2\neq0$), $F_{ij}$ and the spatial metric must further obey the constraint
\be
  b^2h_{ij}=(D-2)F_{ik}F_{jl}h^{lk} ,
 \label{constrF_gen} 
\ee
which means that the base space is {\em almost-Hermitian} \cite{Yanobook_complex} (in addition to being Einstein)\footnote{While this condition is identically satisfied in the case $D=4$, it restricts considerably the permitted spatial geometries when $D>4$ -- cf. section~\ref{sec_static} for more comments and related references. From a geometric viewpoint, it is also worth pointing out that such spacetimes with the 2-form $\bF$ naturally define almost-Robinson manifolds \cite{NurTra02,Trautman02a,Taghavi-Chabert14,FinLeiTag21}.} and $D$ must therefore be {\em even}. Vice versa, when $D$ is {\em odd} (recalling that in this case $F$ must be non-negative, cf. section~\ref{subsec_theory}) one has necessarily $ b^2=0=2 e^2$, i.e., $F=0$, contradicting the assumption made above that the field is non-stealth -- odd dimensional solutions cannot therefore occur here (but they are contained in the discussion of section~\ref{sec_stealth}).

All the above equations hold also in the limit $D=4$, except for \eqref{mu_u} (see section~\ref{sec_integr} and \cite{RobTra62,Stephanibook,OrtPodZof08} for more details on the $D=4$ case).\footnote{Beware of two typos in \cite{OrtPodZof08}: the RHS of (B.13) should read $8P^2(Q_{,1}\xi_1 + Q_{,2}\xi_2)$, while on the RHS of (B.14) there should be a factor 8 (instead of 4).} For $D>4$, it should also be observed that several features of the obtained solutions contrast with those of the linear theory \cite{OrtPodZof08}. First, the non-linearities cause the magnetic term in \eqref{H} to fall off more quickly at infinity and thus make the geometry better behaved (for a slower fall-off as in the linear theory \cite{OrtPodZof08} cf. also related comments in \cite{GibTow06}). They also give rise to a dimensional-independent fall-off of the electric component of \eqref{F_summ} (as already observed in \cite{HasMar07} and in contrast to the standard Coulomb field of the linear theory \cite{Tangherlini63}, see also \cite{OrtPodZof08}). In addition, apart from the electric and magnetic components $e$ and $F_{ij}$, the electromagnetic field \eqref{F_summ} may contain also a radiative null term $F_{ui}$, which is related to a possible mass loss (or gain) encapsulated in eq.~\eqref{mu_u}. This has also to do with the fact that the line-element is in general time-dependent (cf. also footnote~\ref{footn_stat} in section~\ref{sec_static}).\footnote{A similar behaviour in a linear theory is possible provided one considers, instead of a 2-form, a $p$-form in $D=2p$ dimensions \cite{OrtPodZof15}.}

Particular specializations of the above family of solutions may thus describe various physical configurations, such as static dyonic black holes, but also time-dependent solutions with a radiating term in $\bF$. These are discussed, respectively, in sections~\ref{sec_static} and \ref{sec_radiating}.

\section{Ansatz and integration of the field equations}

\label{sec_integr}

\subsection{Robinson-Trautman geometry with an aligned 2-form field}

\label{sec_geom}

A $D$-dimensional Robinson-Trautman spacetime \cite{PodOrt06} is defined by admitting a non-twisting, non-shearing, expanding geodesic null vector field $\bk$. This can be expressed invariantly as \cite{RobTra83}
\be
	k_\mu k^\mu=0 , \qquad k_{[\rho}k_{\mu;\nu]}=0 , \qquad \pounds_{\bk} g_{\mu\nu}=\rho g_{\mu\nu}+k_\mu\xi_\nu+\xi_\mu k_\nu , \qquad \rho\neq 0 . 
	\label{RT_def}
\ee	
The latter condition is precisely the one which defines a non-zero expansion (the case $\rho=0$ corresponding, instead, to Kundt spacetimes \cite{PodOrt06}).

The associated Robinson-Trautman line-element was obtained in adapted coordinates in \cite{PodOrt06}.

It is the purpose of the present paper to determine Robinson-Trautman solutions of the theory~\eqref{action}. We shall restrict to the case of electromagnetic fields that are aligned with $\bk$ \cite{OrtPodZof08}, i.e., 
\be
 F_{\alpha \beta}k^{\beta} = N k_{\alpha} .
 \label{aligned}
\ee

Thanks to \eqref{aligned}, Einstein's equations~\eqref{Einst} reveal that the Ricci tensor is necessarily doubly aligned with $\bk$ (i.e., $R_{\mu\nu}k^\nu\propto k_\mu$). Similarly as in \cite{OrtPodZof15}, this enables one to specialize the form of the general Robinson-Trautman metric obtained in \cite{PodOrt06}, i.e., we can already start from the simplified line-element 
\beq
  & & \d s^2=r^{2}h_{ij}(u,x)\left(\d x^i+ W^{i}\d u\right)\left(\d x^j+ W^{j}\d u\right){-}2\,\d u\d r-2H\d u^2 , \label{geo_metric_text} \\
	& & W^{i}=\alpha^i(u,x)+r^{1-D}{\beta}^i(u,x) \label{W} ,
\eeq
where the function $H$ can depend on all spacetime coordinates, and $h_{ij}$ denotes a (so far unspecified) Riemannian metric in $D-2$ dimensions. For later purposes let us note that
\begin{equation}\label{determinants}
  \sqrt{-g} = r^{D-2}\,\sqrt h ,
\end{equation}
where $g\equiv \det g_{\mu\nu}$.

Using the above coordinates one has
\be
  k^\mu\pa_\mu=\pa_r , \qquad {k_\mu\d x^\mu={-}\d u } ,
\label{k}
\ee
such that $r$ is an affine parameter along $\bk$. Condition~\eqref{aligned} reads
\be
	F_{r i}=0 , \qquad F_{r u}=N .
	\label{F_aligned}
\ee

\subsection{Integration of the field equations}

Eqs.~\eqref{geo_metric_text}, \eqref{W} and \eqref{F_aligned} already ensure that $E_{rr}=0=E_{ri}$ (cf. \cite{PodOrt06,OrtPodZof08,OrtPodZof15} for related discussions).

With \eqref{F_aligned}, eq.~\eqref{Max2} reads
\beq
 & & F_{ij,r}=0, \label{F_ijr}\\
 & & F_{ui,r}=- N_{,i}, \label{F_uir}\\
 & & F_{ij,u}= F_{uj,i} - F_{ui,j} , \label{F_iju}\\
 & & F_{[ij,k]}= 0 , \label{F_ijk} 
\eeq
while eq.~\eqref{Max1} becomes (using also \eqref{geo_metric_text} -- the explicit relation between covariant and contravariant components of $\bF$ can be found in \cite{OrtPodZof08}) 
\beq
	& & \left(r^{D-2}F^{\frac{D}{4}-1}N \right)_{, r}=0, \label{FirstDynamic} \\
	& & \sqrt{h}\left(r^{D-2}F^{\frac{D}{4}-1}F^{ir} \right)_{, r}=-r^{D-2}\left(\sqrt{h}F^{\frac{D}{4}-1}F^{ij} \right)_{, j}, \label{SecondDynamic} \\
	& & \left(\sqrt{h}F^{\frac{D}{4}-1}F^{ir} \right)_{, i}=-\left(\sqrt{h}F^{\frac{D}{4}-1}N\right)_{, u}. \label{ThirdDynamic}
\eeq

The $r$-dependence of $\bF$ is determined by \eqref{F_ijr}, \eqref{FirstDynamic} and \eqref{F_uir}, namely 
\beq
 & & F_{ij}=F_{ij}(u, x) , \label{Fij} \\ 
 & & N=\dfrac{ e(u,x)}{r^2} , \label{Fru} \\ 
 & & F_{ui}=\dfrac{ e_{,i}}{r}-\xi_i(x, u) , \label{Fui} 
\eeq
where $e$ and $\xi_i$ are integration functions.

Substituting \eqref{Fui} into  \eqref{F_iju} gives
\be
 F_{ij,u}=\xi_{i,j}-\xi_{j,i} .
 \label{F_ij,u}
\ee

Eq.~\eqref{F_ijk} simply means that $F_{ij}$ defines a closed 2-form in the spatial base manifold. Consequences of the remaining equations~\eqref{SecondDynamic} and \eqref{ThirdDynamic} will be discussed more easily after \eqref{f=0=e} is obtained below. 

At this stage, the invariant \eqref{F2} (useful in the following) takes the form
\begin{align}
F=r^{-4}F_0 , 
 \label{F_invariant}
\end{align}
where we have defined $F_0$ as in \eqref{F_spat}.

Using \eqref{F_aligned}, \eqref{Fij} and \eqref{F_invariant} one finds that the $(ij)$ component of the RHS of \eqref{Einst} is proportional to $r^{2-D}$. Comparing this with the Ricci tensor component \eqref{Rij} implies ${\beta}^i=0$. Furthermore, a coordinate transformation enables one to set (at least locally) $\alpha^i=0$ \cite{PodOrt06}. From now on we shall thus have in \eqref{geo_metric_text}
\be
 W^i=0 , \label{f=0=e}
\ee
which simplifies several quantities. 

Eq.~\eqref{f=0=e} with \eqref{Fru}, \eqref{Fui}, \eqref{F_invariant} enables one to easily write the remaining Maxwell equations~\eqref{SecondDynamic} and \eqref{ThirdDynamic} as 
\be
 F_0^{\frac{D}{4}-1}\sqrt{h}h^{ij} e_{,j}  = \Big(F_0^{\frac{D}{4}-1} \sqrt{h}  h^{ik} h^{jl} F_{kl} \Big)_{,j} , \qquad 
\Big(F_0^{\frac{D}{4}-1}\sqrt{h}h^{ij} \xi_j \Big)_{,i} =  \Big(F_0^{\frac{D}{4}-1} \sqrt{h}   e\Big)_{,u} .
\ee
Alternatively, these can be rewritten covariantly as $F_0^{\frac{D}{4}-1} e_{,i}  = \big(F_0^{\frac{D}{4}-1}F_{ij}\big)^{||j}$ and $\sqrt{h}(F_0^{\frac{D}{4}-1}h^{ij} \xi_j)_{||i} =(F_0^{\frac{D}{4}-1} \sqrt{h}   e)_{,u}$, where a double bar denotes a covariant derivative in the base space.

Further analysis of various powers of $r$ appearing in $E_{ij}$ determines the $r$-dependence of $H$
\beq
 2H=& & \frac{\mathcal{R}}{(D-2)(D-3)}+\frac{2}{D-2}\big(\ln \sqrt{h}\big)_{,u}r -\frac{2\Lambda}{(D-2)(D-1)}r^2 - \frac{\mu}{r^{D-3}}  \nonumber \\ 
	& & {}+\frac{ { 2\kappa \beta}}{D-2} F_0^{\frac{D}{4}-1} \frac{ b^2 + (D-2)  e^2}{r^{D-2}} , \label{H1}
\eeq
and additionally gives the following conditions
\beq
 & & \R_{ij}=\frac{\R}{D-2}h_{ij}   , \label{constrijr} \\
 & & h_{ij,u} =\frac{2(\ln\sqrt{h})_{,u}}{D-2}h_{ij} , \label{constrijs} \\
 & &  b^2h_{ij}=(D-2)F_{ik}F_{jl}h^{lk} , \label{constrF} 
\eeq
where the identity $h^{ij}h_{ij,u}=2(\ln \sqrt{h})_{,u}$ has been used. Here ${\R_{ij}}$ is the Ricci tensor associated with the spatial metric $h_{ij}$, $\R=h^{ij}\R_{ij}$ its Ricci scalar, and $\mu$ an integration function independent of $r$.

Using \eqref{H1} and \eqref{Rur} one finds that the equation $E_{ur}$ is satisfied identically.

Using \eqref{H1}, \eqref{constrijs} and \eqref{Rui}, the vanishing of the coefficient of the term $r^{2-D}$ in the equation $E_{ui}$ requires
\be
 \mu_{,i} =  {2\kappa \beta} D F_0^{\frac{D}{4}-1}\big(  e \xi_i - F_{ik} \xi_j h^{kj} \big) .
\ee
Coefficients of some other powers of $r$ vanish identically as a consequence of \eqref{idi}.

Finally, with \eqref{H1} and \eqref{Ruu}, the vanishing of the coefficient of the term $r^{2-D}$ in the equation $E_{uu}$ gives
\be
 (D-2)\mu_{,u} = -(D-1) \mu \big( \ln \sqrt{h} \big)_{,u}-  {2\kappa D \beta}  F_0^{\frac{D}{4}-1}h^{ij}\xi_i \xi_j  \qquad (D>4) ,
 \label{Euu}
\ee
while other powers of $r$ in $E_{uu}$ vanish identically as a consequence of \eqref{id0}. For later purposes it is useful to point out that one of those identities reads
\be
 \triangle\left[ F_0^{\frac{D}{4}-1} \left[ b^2 + (D-2) e^2 \right]\right] = D (D-2) F_0^{\frac{D}{4}-1}h^{ij}e_{,i}e_{,j} ,
 \label{id_uu}
\ee
where $\triangle\equiv\frac{1}{\sqrt{h}}\pa_j(\sqrt{h}h^{ij}\pa_i)$ is the Laplace operator in the $(D-2)$-dimensional space with metric $h_{ij}$.

All the equations obtained previously hold also for $D=4$, except for \eqref{Euu}, which applies only to the case $D>4$ -- the reason for this is that $E_{uu}$ contains terms with powers $r^{-2}$ and $r^{2-D}$, which combine precisely when $D=4$ (resulting in an additional term proportional to $\triangle\R$ in \eqref{Euu}, cf. \cite{RobTra62,Stephanibook} and appendix~B of \cite{OrtPodZof08}). On the other hand, for $D=4$ eq.~\eqref{constrijr} is an identity, whereas for $D>4$ it means that the metric $h_{ij}$ is Einstein and therefore $\R=\R(u)$. 
Eq.~(\ref{constrijs}) means that $h_{ij}$ can depend on $u$ only via a conformal factor, i.e., $h_{ij}=h^{1/(D-2)}\,\gamma_{ij}(x)$ \cite{PodOrt06}. 

To conclude, we note that a transformation of the form \cite{RobTra62,PodOrt06} 
\be
 u=u(\tilde u), \qquad r=\tilde r/\dot u(\tilde u) ,
\label{tranf}
\ee
can be used to rescale the first term in \eqref{H1} arbitrarily (provided $\mu$, $e$ and $b$ are also appropriately redefined), so that without losing generality we can hereafter assume $\mathcal{R}=0,\pm (D-2)(D-3)$. Combining all the results obtained above one arrives at the summary given in section~\ref{subsec_summary}.

\section{Static black holes}

\label{sec_static}

\subsection{The solutions}

Here we assume that $\pa_u$ is a Killing vector field of the metric \eqref{ds_generic}, so that the spacetime is static in regions where $H>0$. This requires $h_{,u}=0$, so that $h_{ij}=h_{ij}(x)$, along with $\mu_{,u}=0$ and $Q_{,u}=0$.
Plugging these conditions into \eqref{mu_u}, \eqref{mu_i}, \eqref{Fij,u} one easily concludes that $\xi_i=0$, $\mu=$const, $F_{ij,u}=0$ (and therefore also $b_{,u}=0$) and $ e_{,u}=0$.\footnote{In other words, the time-dependence of the general line-element~\eqref{ds_generic}, \eqref{H} is due to the radiative component of $\bF$ in \eqref{F_summ} (see in particular eq.~\eqref{mu_u}). From a complementary viewpoint, when $\mu\neq0$ staticity of the metric also follows by assuming $\xi_i=0$ in \eqref{F_summ} (and using \eqref{tranf}, see a related discussion in \cite{OrtPodZof08,OrtPodZof15}).\label{footn_stat}}

Now, let us recall the identity \eqref{id_uu}. Since its RHS is non-negative, Hopf's theorem (see, e.g., \cite{Yanobook_complex}) implies that when the base manifold is {\em compact} then both $F_0^{\frac{D}{4}-1} \left[ b^2 + (D-2) e^2 \right]$  and $ e$ must be constant, and therefore such must be also $b$. Having in mind primarily black hole spacetimes, hereafter we shall thus restrict ourselves to the case $ e=$const and $b=$const.

The solutions of interest are thus given by the line-element~\eqref{ds_generic}\footnote{Standard Schwarzschild coordinates are obtained by the well-known transformation ${\d u=\d t-\d r/2H}$ (cf., e.g., \cite{Stephanibook,OrtPodZof08}).} with the base space metric $h_{ij}(x)$ being Einstein, and
\be
 2H=K-\lambda r^2 - \frac{\mu}{r^{D-3}} + \frac{Q^2}{r^{D-2}} ,
 \label{Hstat}
\ee
while the electromagnetic field reads
\be
 \bF = \dfrac{ e}{r^2}\d r \wedge\d u + \dfrac{1}{2}F_{ij}(x)\d x^i \wedge \d x^j ,
 \label{Fstat}
\ee
with \eqref{Q}, \eqref{F_spat}. The constant $K=0,\pm1$ represents the sign of the Ricci scalar of $h_{ij}$, $\mu$ is a mass parameter, while $e$ and $b$ parametrize the electric and magnetic components of $\bF$.\footnote{At least for asymptotically flat purely electric solutions, the mass and electric charge were computed in \cite{HasMar07} using a reduced Hamiltonian action and, perhaps not surprisingly, are determined in terms of the parameters $\mu$ and $e$. The thermodynamics was studied subsequently in \cite{GonHasMar09}. For a definition of the electric and magnetic parts of an arbitrary tensor in any dimension cf., e.g., \cite{Senovilla00,Senovilla01,HerOrtWyl13}.\label{footn_charge}} In the above expressions, the main differences w.r.t. the $D>4$ Einstein-Maxwell solutions \cite{OrtPodZof08} are the better behaved magnetic term in \eqref{Hstat} (the electric and magnetic terms in \cite{OrtPodZof08} fall off as $1/r^{2(D-3)}$ and $1/r^2$, respectively) and the fact that the fall-off of electric field component in \eqref{Fstat} does not depend on $D$ (as noticed in \cite{HasMar07}).

The only remaining field equations reduce to (cf. \eqref{Fij,u}, \eqref{grad_alpha})
\be
 F_{[ij,k]}= 0 , \qquad \Big(\sqrt{h}  h^{ik} h^{jl} F_{kl} \Big)_{,j}=0 , \\
\ee
along with \eqref{constrF_gen}. This means that the 2-form $F_{ij}$ must be closed and coclosed in the base space geometry and that, when $F_{ij}\neq0$, the base space must be {\em almost-K\"ahler} \cite{Yanobook_complex} (and not just almost-Hermitian, as in the general case of section~\ref{subsec_summary}) -- in particular, the only permitted space of constant curvature is flat space \cite{Yanobook_complex}, in which case $F_{ij||k}=0$ and a solution can be easily found in closed form \cite{OrtPodZof08} and interpreted as a black brane (as done in \cite{DHoKra09} in the $D$-dimensional Einstein-Maxwell theory).\footnote{Recall that, apart from flat space, examples of Einstein-K\"ahler spaces are provided by direct products of identical 2-dimensional spaces of constant curvature $S^2\times S^2\times\ldots$ or $H^2\times H^2\times\ldots$, or the complex projective space $\mathbb{C}P^{\frac{n}{2}}$ and the complex hyperbolic space $H_{\mathbb{C}}^{\frac{n}{2}}$ with the Fubini-Study metric \cite{KobNom2}. A thorough description and more examples can be found, e.g., in \cite{Yanobook_complex,KobNom2,Bessebook} (see \cite{Armstrong02,ApoDra03} and references therein for almost-K\" ahler Einstein manifolds). The relevance of these geometries in the context of higher-dimensional charged black hole spacetimes has been already pointed out for linear electrodynamics in \cite{OrtPodZof08,BarCalCha12,OrtPodZof15,OhaNoz15} and for modified theories in \cite{HerOrt20_2}.} This also implies that dyonic (or purely magnetic) solutions cannot be asymptotically flat. By contrast, in the purely electric case ($F_{ij}=0$) the base manifold can be any Einstein space, so in particular a round sphere. The latter solutions were found in \cite{HasMar07} in the case $\Lambda=0$. When $e=0=b$, i.e., $\bF=0$, eq.~\eqref{Hstat} describes familiar Schwarzschild-like black holes of vacuum Einstein's gravity \cite{Tangherlini63,GibWil87,Birmingham99} (see also section~\ref{sec_stealth}).

It is interesting to observe that all the above black hole solutions turn out to be ``immune'' to some corrections to the gravity part of the action~\eqref{action}, i.e., they coincide with electric \cite{Sheykhi12} and magnetic \cite{HerOrt20_2} solutions of certain $f(R)$ gravities. This is no longer true for extensions of \eqref{action} to Gauss-Bonnet gravity, however exact solutions thereof are also known in the case of electric \cite{Hendi09} and dyonic \cite{HerOrt20_2} fields (for Gauss-Bonnet magnetic black holes coupled to different power-like electrodynamics see \cite{MaeHasMar10}).

It is also worth remarking that when $\Lambda=0$ metric~\eqref{Hstat} can also be seen as a vacuum (non-Einstein) solution of pure $R^2$-gravity (in which case $Q^2$ in \eqref{Hstat} is simply an integration constant) \cite{Hendi10,HerOrt20}. When $\Lambda$ and Einstein terms are added to the action, or for more general $f(R)$ theories, this remains true provided the parameters of the theory are suitably fine-tuned \cite{Hendi10,HenEslMou12,HerOrt20}. That this is the case can be traced back to the fact that the Ricci scalar of \eqref{Hstat} is constant \cite{Hendi10,HenEslMou12,HerOrt20,HerOrt20_2}. Similarly, the same metric also solves Einstein's gravity coupled to a conformal scalar field \cite{Giribetetal14}.

\subsection{Horizons and spacetime structure}

Since the vacuum case is known, in the following discussion we shall assume $\bF\neq0$. As mentioned in section~\ref{subsec_summary}, there is a timelike curvature singularity at $r=0$ (similarly as in the {R}eissner--{N}ordstr{\"{o}}m spacetime, cf., e.g., \cite{GriPodbook} and references therein). We can therefore restrict ourselves to the range $r>0$. 

\begin{figure}
 \includegraphics[width=9cm]{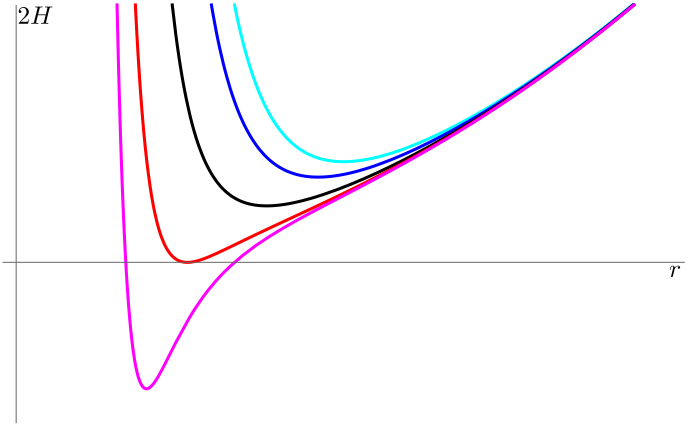}
 \caption{Plot of the function $2H(r)$ (eq.~\eqref{Hstat}) for $D=8$ in the case $\Lambda<0$, $K=0$. For the cosmological constant and the electromagnetic field strength we have chosen the values $\lambda=-1$ (recall \eqref{Q}), $Q^2=1$, while the mass parameter $\mu$ ranges from 2 (lower, magenta curve) to $-6$ (upper, light blue curve). Intersections with the $r$-axis represent Killing horizons. In particular, the red curve denotes a double horizon (at $\mu\approx 1.458$, within numerical accuracy).}
 \label{fig_Lambda<0_K=0}
\end{figure}

\begin{figure}
 \includegraphics[width=9cm]{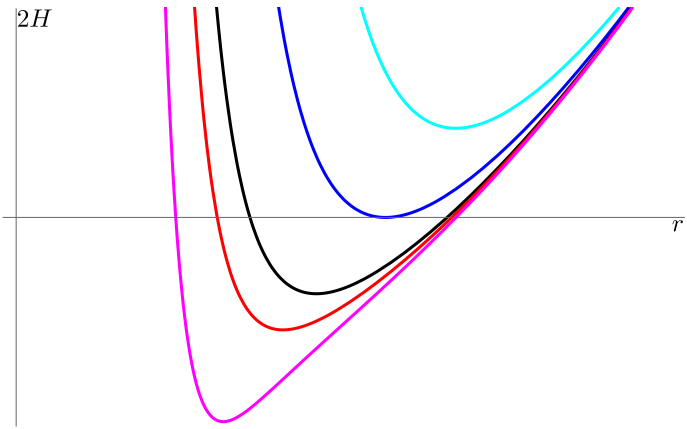}
  \caption{Plot of the function $2H(r)$ for $D=8$ in the case $\Lambda<0$, $K=-1$. Here $\lambda=-0.7$, $Q^2=0.05$, and $\mu$ ranges from 0.1 (lower, magenta curve) to $-1$ (upper, light blue curve). The dark blue curve denotes a double horizon (at $\mu\approx -0.251$, within numerical accuracy).}
 \label{fig_Lambda<0_K<0}
\end{figure}

\begin{figure}
 \includegraphics[width=9cm]{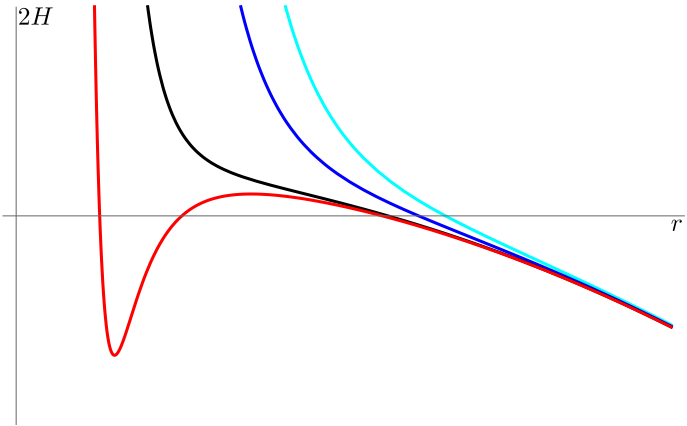}
  \caption{Plot of the function $2H(r)$ for $D=8$ in the case $\Lambda>0$, $K=1$. Here $\lambda=0.5$, $Q^2=0.05$, and $\mu$ ranges from 0.16 (lower, red curve), over 0 (black curve), to $-4$ (upper, light blue curve). The red curve displays a case when there are three distinct Killing horizons (inner, black hole and cosmological horizons).}
 \label{fig_Lambda>0_K>0}
\end{figure}

\begin{figure}
 \includegraphics[width=9cm]{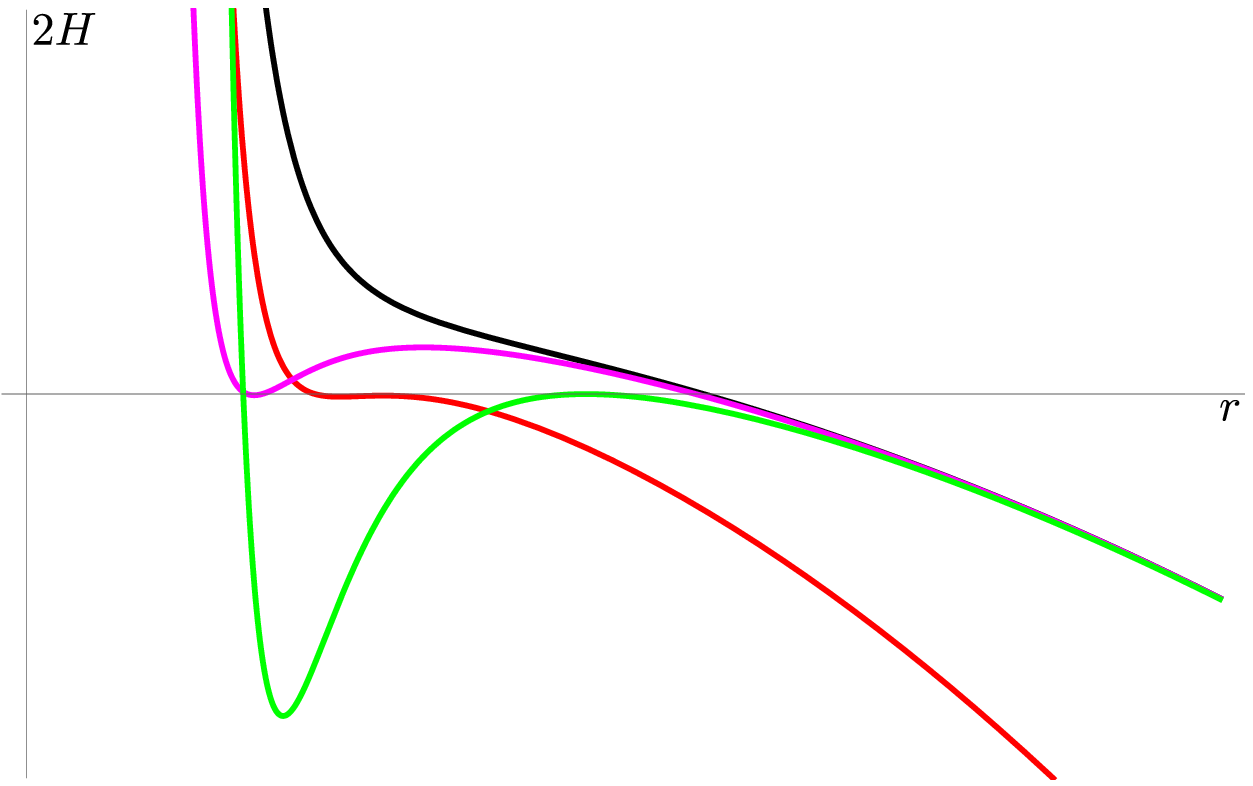}
  \caption{Plot of the function $2H(r)$ for $D=8$ in the case $\Lambda>0$, $K=1$. Note that in this case we have chosen different values of the parameters for each curve -- this is just a technicality that enables us to plot in a single figure various curves describing different special cases. Namely, the magenta (($Q^2, \lambda, \mu) \approx (0.05, 0.5, 0.127)$) and green ($(Q^2, \lambda, \mu) \approx (0.5, 0.5, 1.119)$) curves contain a double horizon, while the red curve ($(Q^2, \lambda, \mu) \approx (0.15, 1.08, 0.293)$) denotes a triple one. The black curve ($(Q^2, \lambda, \mu) = (0.05, 0.5, 0)$) possesses only a single horizon.}
 \label{fig_Lambda>0_K>0_2}
\end{figure}

The spacetime is static in regions where $H>0$, whereas positive values of $r$ such that $H=0$ represent Killing horizons. The latter are defined by positive real roots of the polynomial
\be
 -\lambda r^D+Kr^{D-2}  - \mu r+Q^2=0 ,
 \label{polynomial}
\ee 
where the last term is positive by the conditions on $\beta$ (cf. section~\ref{sec_intro}). Using Descartes's rule of signs one can place restrictions on the signs of the parameters $\Lambda$, $K$ and $\mu$ in order for such roots to exist, and simultaneously count how many of those may occur. This is straightforward and summarized in table~\ref{table_horiz}. Note that the counting of the roots includes their multiplicities, so the case of 2 roots also possibly includes a double one (an extreme horizon), while the case of 3 roots (in general corresponding to a cosmological, an outer and an inner horizon) also allows for a single root accompanied by a double one, or for a triple root. In particular, the latter case occurs at $r=r_3\equiv\sqrt{(D-3) (D-2)^2}/\sqrt{2D\Lambda}$ when $\Lambda>0$, $K=1$, and for the special values $\mu=\frac{4 D \Lambda  r_3^{D-1}}{(D-3) (D-2) (D-1)}$, $(D-2)^2Q^2=4 \Lambda  r_3^D$ (cf. \cite{Lake79} in the case $D=4$). More generally, assessing the precise number of horizons which do actually occur (e.g., 2 vs. 0, or 3 vs. 1) depends on the particular chosen ranges of the parameters. This is illustrated in figures \ref{fig_Lambda<0_K=0}--\ref{fig_Lambda>0_K>0_2} by plotting $H$ as a function of $r$ for various values of $\mu$ in the exemplificative cases $\Lambda<0$ with $K=0$ and $K=-1$, and $\Lambda>0$ with $K=1$.  All combinations of signs which do not appear in table~\ref{table_horiz} (for example, $(\Lambda\le0,K\ge0,\mu\le0)$) describe naked singularities. We observe that the structure of Killing horizons is qualitatively similar to the one of the four-dimensional (A)dS-Reissner-Nordstr{\"o}m metrics \cite{Lake79,GriPodbook}. Some related comments and plots complementary to the one given above can be found in \cite{Hendi10} for the case $\Lambda=0$ and in \cite{Sheykhi12} for $\Lambda<0$ with $K=1$.

\begin{table}[t]
\begin{center}
\begin{tabular}{|c||c|c|c|c|} 
  \hline 
 $\Lambda$ & \quad $K$ \quad  & \quad $\mu$ \quad & {$\#$ horizons} & {$\pa_u$ for $r\to\infty$}  \\ \hline\hline   
  $<0$ & $\ge0$  & $>0$  & 2,0 &  timelike \\ 
		   & $<0$ & any  & 2,0 &  \\ \hline 
			 & $>0$ & $>0$ & 2,0 & timelike \\ 
	$0$	 & $0$ & $>0$ & 1 & spacelike \\ 
			 & $<0$ & any & 1 & spacelike \\ \hline 				
			 & $>0$ & $>0$  & 3,1 &  \\ 
	$>0$ & $>0$ & $\le 0$  & 1 &  spacelike \\ 
			 & $\le0$ & any & 1 &  \\ \hline  						
\end{tabular} 
\end{center}
		\caption{Possible combinations of the signs of the parameters $\Lambda$, $K(=0,\pm1)$ and $\mu$ that permit the polynomial equations~\eqref{polynomial} to admit positive real roots. In the fourth column the possible number of roots is indicated (including their multiplicities) for each case (in some cases there may be more than one possibility, depending on the specific range of the parameters). The last column displays the character of the Killing vector field $\pa_u$ sufficiently close to infinity, i.e., in the outer region. See the main text for more information.}
 \label{table_horiz}
\end{table}

The asymptotic properties at $r\to\infty$ are determined by the sign of $\Lambda$, similarly as in  vacuum Einstein's gravity (cf. \cite{Birmingham99}). In particular, when the base space is a round sphere (which requires $F_{ij}=0$, as mentioned above), the spacetime is asymptotically flat or (A)dS. When $F_{ij}\neq0$ this is not possible, however, the spacetime is asymptotically locally AdS if $h_{ij}$ is a flat metric and $\Lambda<0$. The Killing vector field $\pa_u$ is spacelike near infinity when $\Lambda>0$ and timelike for $\Lambda<0$. When $\Lambda=0$, it is timelike if $K>0$ and spacelike if $K<0$ (if also $K=0$, its character is determined by the sign of $\mu$; if $K=0=\mu$, it is timelike), see also table~\ref{table_horiz}.

\section{Time-dependent solutions}

\label{sec_radiating}

In this section we consider non-stealth solutions with a non-zero radiating term $\xi_i\neq0$ in the electromagnetic field~\eqref{F_summ}.  As discussed in section~\ref{sec_static}, this ensures that $\pa_u$ is not Killing. Eq.~\eqref{mu_i} further implies $\mu_{,i}\neq0$, while eq.~\eqref{mu_u} shows how $\xi_i$ contributes to a mass loss due to electromagnetic radiation as the retarded time $u$ evolves. The electromagnetic energy flux along the Robinson-Trautman null vector field $\bk= \pa_r$ is measured by the leading term of the scalar $T_{\mu\nu}n^\mu n^\nu$, where $\bn=\pa_u-H\pa_r$ is another null vector field such that $n^\mu k_\mu=-1$. One finds ($T_{\mu\nu}$ is defined in section~\ref{subsec_theory})
\be
 T_{\mu\nu}n^\mu n^\nu=\beta D F_0^{D/4-1}\frac{h^{ij}\xi_i\xi_j}{r^{D-2}}+{\cal O}(r^{1-D}) ,
 \label{flux}
\ee
where the leading term actually takes the same value in any frame parallelly transported along $\bk$ and is thus an invariant quantity.  As expected from the above comments, it vanishes iff $\xi_i=0$.

If the base manifold is taken to be compact, the same argument as used in section~\ref{sec_static} gives $e_{,i}=0=b_{,i}$ -- but here $e$ and $b$ can depend on $u$. We observe that in four dimensions there exist simple (Vaidya-like) explicit solutions describing the evaporation of a white hole (or, by time-reversal, black hole formation) by emission (collapse) of electromagnetic radiation, which has the form of a null field \cite{Senovilla15}. However, no analog of those solutions is possible here, both because $\mu_{,i}\neq0$ and because null fields are stealth and thus do not backreact.

Time-dependent solutions can be understood as dynamical extensions of the static black holes of section~\ref{sec_static}. It is an open question whether they indeed do settle down to a static configuration at late times (at least under suitable initial conditions) or whether they develop instabilities as some of their four-dimensional counterparts \cite{LunCho94,KozKreReu08}. This would deserve a separate investigation. However, even if the future evolution was sound (with a possible extension across a future horizon), one may still expect the past evolution to be singular \cite{Chrusciel91}, preventing one from sensibly defining a past event horizon. From this viewpoint it is more appropriate to study, instead, past quasi-local horizons \cite{Tod86,Tod89,LunCho94,ChoLun99,NatTaf08,PodSvi09}. In the following we show how one can do that for Robinson-Trautman spacetimes in higher dimensions, and specifically for solutions of the theory~\eqref{action} (see \cite{Svitek11} for earlier results in the presence of null radiation but without an electromagnetic field).

\subsection{General setup in Robinson-Trautman spacetimes}

Given the spacetime \eqref{ds_generic}, let us consider a family of $(D-2)$-dimensional spacelike surfaces ${\cal S}$ defined by
\be
 u=u_0 , \qquad r=X(x;u_0) , 
 \label{S}
\ee
where $u_0$ is a constant parameter and $X$ a positive function (at this stage arbitrary) of its arguments.

Similarly as in \cite{Svitek11} (see also the earlier work \cite{LunCho94,ChoLun99} in four dimensions), a null frame adapted to the above surfaces is defined by
\be
  \bk= \pa_r , \quad \bn=\pa_u+\frac{1}{2}\left(-2H+r^{-2}h^{ij}Y_{,i}Y_{,j}\right)\pa_r+r^{-2}h^{ij}Y_{,i}\pa_j , \quad \bm_{(\alpha)}=r^{-1}\tilde m_{(\alpha)}^i\left(\pa_i+X_{,i}\right)\pa_r ,
\ee
where $\tilde m_{(\alpha)}^i$ are the components of an orthonormal frame in the base space with metric $\bh$ (i.e., $h_{ij}\tilde m_{(\alpha)}^i\tilde m_{(\beta)}^j=\delta_{(\alpha)(\beta)}$). The null vectors $\bk$ and $\bn$ are, respectively, the outgoing and ingoing future-oriented normals to the surfaces ${\cal S}$, while the $\bm_{(\alpha)}$ span such surfaces. The expansion $\Theta_{\bk}$ of $\bk$ is positive by construction, whereas for $\bn$ one finds
\be
 (D-2)\Theta_{n}=X^{-1}\left[\triangle\ln X-(D-2)H+X(\ln\sqrt{h}),_{u}+\frac{1}{2}(D-4)h^{ij}(\ln X)_{,i}(\ln X)_{,j}\right] ,
 \label{Th_n}
\ee
where $H$ has to be evaluated with \eqref{S} holding.

For choices of $X$ such that $\Theta_{n}>0$, the corresponding ${\cal S}$ are (past) trapped surfaces (at least if they are compact) \cite{Penrose65prl}, while solutions of the equation $\Theta_{n}=0$ (if they exist) represent marginally trapped surfaces \cite{Penrose73}. In the latter case, the $(D-1)$-dimensional hypersurface ${\cal H}$ defined by $r=X(x,u)$ (where now $u$ is not fixed) is thus a marginally trapped tube (foliated by the surfaces ${\cal S}$). Furthermore, if ${\cal H}$ is spacelike, it defines a {\em dynamical horizon} \cite{AshKri02}.\footnote{To be precise, in \cite{AshKri02} a ``time-reversed'' situation is considered, i.e., (future) horizons with $\Theta_{\bk}<0$.}  Since the normal to ${\cal H}$ is $N_\mu\d x^\mu=-\d r+X_{,u}\d u+X_{,i}\d x^i$, this happens when 
\be
 2(H+X_{,u})+h^{ij}(\ln X)_{,i}(\ln X)_{,j}<0 ,
\ee
where we have imposed $r=X$ and also $H$ has to be evaluated at ${\cal H}$.

\subsection{Past horizons in Einstein gravity with conformally invariant electrodynamics}

So far the discussion has been general, i.e., it applies to any spacetime of the form \eqref{ds_generic}. In the case of the solutions constructed in the present paper, $H$ is given by \eqref{H}, so that using \eqref{Th_n} the equation $\Theta_{n}=0$ defining marginally trapped surfaces reads explicitly 
\be
  2\triangle\ln X-(D-2)\left(K-\lambda X^2 - \frac{\mu}{X^{D-3}}+\frac{Q^2}{X^{D-2}}\right)+(D-4)h^{ij}(\ln X)_{,i}(\ln X)_{,j}=0 .
	\label{PT}
\ee

This is a non-linear PDE for the unknown function $X(x;u_0)$, and for $Q=0$ it reduces to a result of \cite{Svitek11}.  As noticed there, the last term in \eqref{PT} makes the non-linearity worse when $D>4$.\footnote{By setting $D=4$ one recovers the equation first obtained in \cite{Tod86,Tod89} for $\Lambda=0=Q$, and extended to more general cases in \cite{LunCho94,PodSvi09,TahSvi16}.} Recall that $\mu$ and $Q$ are generically functions of $(u,x)$ (but $Q_{,i}=0$ if the base space is compact), constrained by the field equations \eqref{Fij,u}--\eqref{mu_u}. As suggested in \cite{Tod89} in four dimensions, solutions to \eqref{PT} define an analog of the past horizon in Robinson-Trautman spacetimes. However, proving existence (and, possibly, uniqueness) of such solutions requires a thorough and rigorous mathematical analysis which goes well beyond the scope of the present paper, and we leave it for future investigations (see \cite{Tod89} for the original results for four dimensional vacua, and \cite{Svitek11} for a modification thereof suitable for the case $D>4$).

\section{Stealth solutions}

\label{sec_stealth}

\subsection{General characterization of stealth solutions}

First of all, let us prove that, as pointed out in section~\ref{sec_intro}, a 2-form $\bF$ has a vanishing energy-momentum tensor in the theory~\eqref{action} iff $F=0$. 
That the latter condition is sufficient follows obviously from \eqref{Einst}. To see that it is also necessary, it suffices to set up an orthonormal frame $\{\mathbf{e}_{\hat0}, \mathbf{e}_{\hat i} \}$ (such that $\hat i, \hat j=1,\ldots,D-1$ and $\mathbf{e}_{\hat0} \cdot \mathbf{e}_{\hat0} = -1$, $\mathbf{e}_{\hat i} \cdot \mathbf{e}_{\hat j} = \delta_{\hat i\hat j}$). Then one sees that the $(\hat0\hat0)$ component of the RHS of \eqref{Einst} vanishes only if either $F=0$ or $(D-2)F_{0\hat i}F_{0\hat i}+F_{\hat i\hat j}F_{\hat i\hat j}=0$ (where we used $F=-2F_{\hat0\hat i}F_{\hat0\hat i}+F_{\hat i\hat j}F_{\hat i\hat j}$). However, the latter expression is non-negative and vanishes iff $F_{\hat0\hat i}=0=F_{\hat i\hat j}$, which is equivalent to the trivial configuration $F_{\mu\nu}=0$. Therefore the only possible stealth fields are those with $F=0$, as we wanted to show. In particular, all null fields are stealth. 

For a stealth field the Maxwell equations~\eqref{Max1} are also identically satisfied and, in order to have a solution, it thus suffices to ensure that $\d\bF=0$. Any closed 2-form $\bF$ hence provides a solution to the theory~\eqref{action} in any Einstein spacetime (other matter fields can obviously be included provided the Einstein equations~\eqref{Einst} are modified accordingly -- yet a stealth $\bF$ will not affect those).\footnote{The fact that null electromagnetic fields may be simultaneous solutions of large classes of electrodynamic theories (and thus have ``universal'' properties) has first been pointed out in \cite{Schroedinger35,Schroedinger43} and investigated more systematically recently \cite{OrtPra16,OrtPra18,HerOrtPra18} (see also \cite{Deser75} for earlier observations).} We also observe that for stealth fields the number of dimensions $D$ can also be odd, since the quantity $F^{D/4}$ is identically zero.

\subsection{Stealth solutions in Robinson-Trautman spacetimes}

Let us now specialize the results outlined above to the case of aligned stealth fields in Robinson-Trautman spacetimes. Since the field is stealth, the Robinson-Trautman metric must be Einstein and thus \cite{PodOrt06} of the form \eqref{ds_generic}, where the spatial metric $h_{ij}=h^{1/(D-2)}(u,x)\gamma_{ij}(x)$ is also Einstein (more details below). The alignment assumption~\eqref{aligned} (i.e., \eqref{F_aligned}) together with the conditions $F=0$ (stealth) and $\d\bF=0$ (closed) mean that $\bF$ can be written as
\be
 \bF = \dfrac{e}{r^2}\d r \wedge\d u + \left(\dfrac{e_{,i}}{r}-\xi_i \right)\d u \wedge \d x^i + \dfrac{1}{2}F_{ij}\d x^i \wedge \d x^j ,
 \label{F_stealth}
\ee
where the functions $e$, $\xi_i$ and $F_{ij}$ depend on $(u,x)$, and further obey
\be
  2e^2=F_{ik} F_{jl} h^{ij} h^{kl} , \qquad F_{[ij,k]}= 0 , \qquad F_{ij,u}=\xi_{i,j}-\xi_{j,i} .
\ee

Simple examples are given by a generalized Coulomb field with $e_{,i}=\xi_i=F_{ij}=0$, or by null fields with $e=F_{ij}=0$ and $\xi_i=\varphi_{,i}$  (where $\varphi(u,x)$ is an arbitrary real function).

Concerning the possible background geometries, let us recall that vacuum Robinson-Trautman spacetimes in Einstein gravity consist of two subclasses \cite{PodOrt06}. Generically one can arrive at a canonical form of the metric \eqref{ds_generic} with $h_{,u}=0$ and 
\be
 2H=K-\lambda r^2-\frac{\mu}{r^{D-3}} \qquad (K=0,\pm 1) ,
\ee
where $\mu$ and $K$ (such that $\R=K(D-2)(D-3)$) are constants. The base space metric $h_{ij}(x)$ can be any Einstein space. These spacetimes describe generalized Schwarzschild black holes \cite{Tangherlini63,GibWil87,Birmingham99}.

In the special case $\mu=0$, the $u$-dependence of the spatial metric cannot in general be removed \cite{PodOrt06,Ortaggio07}, and one has instead 
\be
 2H=K+\frac{2}{D-2}\big(\ln \sqrt{h}\big)_{,u}r -\lambda r^2  \qquad (K=0,\pm 1) .
\ee
The base space metric $h_{ij}(u,x)$ is Einstein and further constrained \cite{Ortaggio07} by being conformal to other Einstein spaces, and thus belongs to the class studied thoroughly in \cite{Brinkmann24,Brinkmann25}.

\section{Conclusions}

\label{sec_concl}

We have presented the complete family of Robinson-Trautman spacetimes admitting an aligned conformally invariant electromagnetic field in the $D$-dimensional theory~\eqref{action} put forward in \cite{HasMar07}. The main differences w.r.t. the linear theory studied in \cite{OrtPodZof08} include a better behaved magnetic term in the metric function $H$, the existence of radiative solutions and the possibility of stealth fields.

A subclass of these metrics represents static dyonic black holes/branes which generalize in various ways an earlier purely electric solution of \cite{HasMar07}. In particular, the presence of a magnetic field allows also for even dimensions which are not a multiple of four. On the other hand, it constrains the horizon geometries more severely (eq.~\eqref{constrF_gen}), in particular ruling out asymptotically flat solutions. Various properties of the general class, such as the structure of horizons, have been clarified.

Furthermore, a new branch of solutions includes time-dependent spacetimes. These describe dynamical black holes emitting (or receiving) electromagnetic radiation. In such a context, quasi-local characterizations of horizon are useful in clarifying geometric properties. We have commented on marginally trapped surfaces and the equation which defines a possible family of past horizons (in the sense of \cite{Tod89}). Further study will be required to assess the existence and uniqueness properties of the solutions of such equation. Another interesting open question concerns the stability of the time evolution of Robinson-Trautman spacetimes. We only remark here that (as already noticed in the vacuum case \cite{PodOrt06}) the $D>4$ ``Robinson-Trautman equation''~\eqref{Euu} presents qualitatively new features as opposed to the well-studied $D=4$ case (notably missing the term corresponding to a Calabi flow when $D=4$ \cite{Tod89}), and one may thus possibly expect a significantly different behaviour in higher dimensions. 

Some ancillary results have been presented in the appendices. In particular, we deem the conclusions of appendix~\ref{app_conserv} to be of some interest in their own right. Since they are theory-independent, they will prove useful to future studies of Robinson-Trautman spacetimes (and in particular of static black holes) also in different contexts. 

Future work may point at extensions of our investigation beyond Einstein's gravity, still in the context of the electrodynamics of \cite{HasMar07}. Some results about static black holes are already available, see \cite{Hendi09,MaeHasMar10,Stetsko19,HerOrt20_2}. Analyzing the thermodynamics of the obtained solutions and their modifications would also be of considerable interest (cf. \cite{GonHasMar09}). More general power-like electrodynamics \cite{HasMar08} are also worth considering, also from the viewpoint of string theory (cf., e.g., \cite{MaeHasMar10}). Not possessing conformal invariance they may display rather different properties (cf. also \cite{HasMar08,HabHal09,MaeHasMar10,Stetsko19}).

\section*{Acknowledgments}

We are grateful to Sourya Ray and Martin \v Zofka for reading the manuscript and for their comments. M.O. was supported by research plan RVO: 67985840 and research grant GA\v CR 19-09659S.

\renewcommand{\thesection}{\Alph{section}}
\setcounter{section}{0}

\renewcommand{\theequation}{{\thesection}\arabic{equation}}
\setcounter{equation}{0}

\section{Conservation equation in Robinson-Trautman spacetimes}

\label{app_conserv}

In this appendix we discuss certain properties of the structure of the equations of motion for general Robinson-Trautman spacetimes in an arbitrary diffeomorphism-invariant metric theory of gravity, including arbitrarily coupled matter fields. Certain results obtained previously (often after tedious computations) in several special cases \cite{PodOrt06,OrtPodZof08,OrtPodZof15} (see \cite{Stephanibook} for the Einstein-Maxwell theory in four dimensions) are thus rederived in a more compact and general way. The present discussion applies, in particular, also to the theory considered in the main body of the paper, thereby allowing one to get rid of a redundancy in the gravity part of the field equations studied there.

Let us consider a diffeomorphism-invariant theory of gravity of the form 
\be
 S=\int\d^Dx\sqrt{-g}{\cal L}(\bR,\nabla\bR,\ldots,\bPsi,\nabla\bPsi,\ldots) ,
\label{action_gen}
\ee
where ${\cal L}$ is a scalar invariant constructed locally from the Riemann tensor $\bR$, the matter fields $\bPsi$ and their covariant derivatives of arbitrary order (following \cite{IyeWal94}, here $\bPsi$ stands for an unspecified collection of tensor fields with arbitrary index structure). 

Extremizing the action w.r.t. $\bg$ produces the gravity part of the corresponding equations of motion $\ensuremath{\boldsymbol{E}}=0$, where $\ensuremath{\boldsymbol{E}}$ is a symmetric 2-tensor defined by \cite{Eddington_book,IyeWal94}
\be
 E_{\mu\nu}\equiv\frac{1}{\sqrt{-g}}\frac{\delta\left(\sqrt{-g}\cal L\right)}{\delta g^{\mu\nu}} . \label{grav_eq} 
\ee

The explicit form of the matter field equations is not relevant to the following discussion. However, once they are satisfied (as we assume hereafter), one obtains that $\bE$ is conserved \cite{Eddington_book,Horndeski74_unp,Horndeski76,Anderson78,IyeWal94}, i.e., 
\be
	\nabla_\nu E^{\mu\nu}=0 .
	\label{Bianchi}
\ee	

For a metric of the form~\eqref{geo_metric_text}\footnote{For the results of this appendix, the particular form of $W^i$ given in \eqref{W} will not be needed -- i.e., they apply to any Robinson-Trautman geometry subject to the only condition $R_{rr}=0$ \cite{PodOrt06}.} one finds \cite{PodSva15} $\Gamma^r_{rr}=\Gamma^u_{rr}=\Gamma^u_{ru}=\Gamma^u_{ri}=\Gamma^i_{rr}=0$ and $\Gamma^i_{rj}=r^{-1}\delta^i_j$, $\Gamma^u_{ij}=r^{-1}g_{ij}$, so that the various components of~\eqref{Bianchi} read
\beq 
 & & \!\!\!\!\!  E^{uu}_{\phantom{uu},u}+E^{ur}_{\phantom{ur},r}+E^{ui}_{\phantom{ui},i}+ (2\Gamma^u_{uu}+\Gamma^r_{ru}+\Gamma^i_{iu})E^{uu}+ (3\Gamma^u_{ui}+\Gamma^r_{ri}+\Gamma^j_{ji})E^{ui}+\Gamma^u_{ij}E^{ij}+\Gamma^i_{ir}E^{ur}=0 , \label{bianchi+1} \\
 & & \!\!\!\!\! E^{iu}_{\phantom{iu},u}+E^{ir}_{\phantom{ir},r}+E^{ij}_{\phantom{ij},j}+ 2(\Gamma^i_{ru}E^{ru}+\Gamma^i_{rj}E^{rj}+\Gamma^i_{uj}E^{uj})+\Gamma^i_{uu}E^{uu}+\Gamma^i_{jk}E^{jk} \nonumber \\
 & & \qquad\qquad\qquad\qquad\qquad\qquad\qquad{}+\Gamma^{\nu}_{\nu u}E^{iu}+  \Gamma^{j}_{j r}E^{ir}+\Gamma^{\nu}_{\nu k}E^{ik} = 0  , \label{bianchi0} \\
 & & \!\!\!\!\! E^{ru}_{\phantom{ru},u}+E^{rr}_{\phantom{rr},r}+E^{ri}_{\phantom{ri},i}+(3\Gamma^r_{ru}+\Gamma^u_{uu}+\Gamma^j_{ju})E^{ru}+(3\Gamma^r_{ri}+\Gamma^u_{ui}+\Gamma^j_{ji})E^{ri} \nonumber \\
& & \qquad\qquad\qquad\qquad\qquad\qquad\qquad{}+\Gamma^r_{uu}E^{uu}+2\Gamma^r_{ui}E^{ui}+\Gamma^r_{ij}E^{ij}+\Gamma^i_{ir}E^{rr} = 0. \label{bianchi-1}
\eeq

Let us now assume the field equations $E^{uu}=0$ and $E^{ui}=0$ have already been solved. Then eq.~\eqref{bianchi+1} reduces to 
\be
 (r^{D-2}E^{ur})_{,r}+r^{D-1}h_{ij}E^{ij}=0 . 
 \label{id1}
\ee
From this condition we learn that the spatial trace of $E^{ij}$ does not provide an equation independent of $E^{ur}$, by virtue of the identity \eqref{Bianchi} (and of the field equations that have already been solved; cf. also \cite{HerOrt20} for related comments in a special case). Alternatively, one can also say that terms contained in $E^{ur}$ that are proportional to powers of $r$ different from $1/r^{D-2}$ necessarily vanish, once $E^{ij}=0$ has been solved.

Once also $E^{ur}=0$ and $E^{ij}=0$ have been solved, eq.~\eqref{bianchi0} becomes
\be
 (r^{D}E^{ir})_{,r}=0 . 
 \label{idi}
\ee
This means that terms of $E^{ir}$ that are proportional to powers of $r$ different from $1/r^{D}$ vanish identically.

Finally, after also $E^{ir}=0$ has been solved, eq.~\eqref{bianchi-1} gives the last identity
\be
 (r^{D-2}E^{rr})_{,r}=0 . 
 \label{id0}
\ee
Therefore, terms of $E^{rr}$ proportional to powers of $r$ different from $1/r^{D-2}$ are zero identically.

The above results are clearly theory-independent, only relying on the form of the metric ansatz~\eqref{geo_metric_text}. As mentioned at the beginning of this appendix, explicit examples of such kind of identities for particular theories have been worked out in \cite{Stephanibook,PodOrt06,OrtPodZof08,OrtPodZof15}.

Let us mention in passing that, similarly as done above, one can also analyze consequences of the generalized Bianchi identity~\eqref{Bianchi} also in the case of Kundt spacetimes, arriving at somewhat different conclusions. This will be discussed elsewhere.

\renewcommand{\theequation}{{\thesection}\arabic{equation}}
\setcounter{equation}{0}

\section{Ricci tensor of Robinson-Trautman spacetimes}

\label{app_Ricci}

Here we follow appendix~A of \cite{OrtPodZof15} (cf. also \cite{PodOrt06,OrtPodZof08,PodSva15}). 

For a metric of the form \eqref{geo_metric_text} one has $R_{rr}=0$ identically. Assuming also \eqref{W}, one further obtains $R_{ri}=0$ and
\beq
 R_{ij}=&&  {\mathcal R_{ij}}-r^{4-D}\left(r^{D-3}2H\right)_{,r}h_{ij}-r^{2(2-D)}\frac{(D-1)^2}{2}h_{ik}h_{jl}  {\beta}^{k}{\beta}^{l} \nonumber\\
 && {}-r\left[\frac{D-2}{2}\Big(2h_{k(i}{{\alpha}^{k}}_{,j)}+{\alpha}^{k}h_{ij,k}{-}h_{ij,u}\Big)
    +\left({\alpha}^{k}_{\;,k}+{\alpha}^{k}(\ln\sqrt{h})_{,k}{-}(\ln\sqrt{h})_{,u}\right)h_{ij}\right]\nonumber\\
 && {}+r^{2-D}\left[\frac{1}{2}\left(2h_{k(i}{{\beta}^{k}}_{,j)}+{\beta}^{k}h_{ij,k}\right)
    -\left({\beta}^{k}_{\;,k}+{\beta}^{k}(\ln\sqrt{h})_{,k}\right)h_{ij}\right] , \label{Rij}
\eeq
where ${\mathcal R_{ij}}$ is the Ricci tensor associated with the spatial metric $h_{ij}$, and a partial derivative w.r.t. $x^j$ is denoted by a comma followed by $j$. 

With the further assumption $W^i=0$ (which is precisely what we need in section~\ref{sec_integr}), the remaining Ricci components take the form
\beq
 R_{ur} & = & r^{2-D}\left(r^{D-2}H_{,r}\right)_{,r}-r^{-1}(\ln\sqrt{h})_{,u}   , \label{Rur} \\
 R_{ui}&=& r^{4-D}\left(r^{D-4}H_{,i}\right)_{,r}
  + \frac{1}{2}\left(h^{jk}h_{ik,u}\right)_{,j} \nonumber\\
 && {}+\frac{1}{2}h^{jk}h_{ik,u}(\ln\sqrt{h})_{,j} -\frac{1}{4}h^{jk}h^{lm}h_{kl,u}h_{jm,i}-(\ln\sqrt{h})_{,ui} ,    \label{Rui} \\
 R_{uu}&= & 2HR_{ur} {-}r^2(r^{-2}H)_{,r}(\ln\sqrt{h})_{,u}
{+}(D-2)r^{-1}H_{,u} \nonumber \\
&& {}+r^{-2}\triangle H -(\ln\sqrt{h})_{,uu}-\frac{1}{4}h^{il}h^{jk}\,h_{ij,u}h_{kl,u}      , \label{Ruu}
\eeq
where $\triangle$ is the transverse Laplace operator as in \eqref{id_uu}.

\renewcommand{\theequation}{{\thesection}\arabic{equation}}
\setcounter{equation}{0}

\providecommand{\href}[2]{#2}\begingroup\raggedright\endgroup


\end{document}